\documentclass{emulateapj}

\def\msun{\mbox{M$_\odot$}}

\def\kms{{\rm\,km\,s^{-1}}}
\def\kmskpc{{\rm\,km\,s^{-1}{kpc}^{-1}}}

\def\Myr{{\rm\,Myr}}
\def\Gyr{{\rm\,Gyr}}
\def\deg{{^\circ}}

\def\kpc{{\rm\,kpc}}
\def\pc{{\rm\,pc}}

\def\mathnew{\mathsurround=0pt}   
\def\simov#1#2{\lower .5pt\vbox{\baselineskip0pt  
    \lineskip-.5pt\ialign{$\mathnew#1\hfil##\hfil$\crcr#2\crcr\sim\crcr}}}

\def\'#1{\ifx#1i{\accent"13\i}\else{\accent"13#1}\fi}

\begin{document}    
\slugcomment{{\em Published in Astrophysical Journal Letters}}         
\shorttitle{Kinematic Constraints on Galactic Structure}     
\shortauthors{Antoja et al. 2009}  

\title{Stellar Kinematic Constraints on Galactic Structure Models Revisited: Bar and Spiral Arm Resonances}

\author{T. Antoja$^{1}$, O. Valenzuela$^{2}$, B. Pichardo$^{2}$, E. Moreno$^{2}$, F. Figueras$^{1}$, D. Fern\'andez$^{1}$ }        

 \affil{$^1$ Institut de Ci\`encies del Cosmos, IEEC-UB, Universitat de Barcelona, Mart\'i i Franqu\`es, 1, E-08028 Barcelona, Spain; tantoja@am.ub.es
\\ $^2$ Instituto de Astronom\'ia, Universidad Nacional
Aut\'onoma de M\'exico, A.P. 70-264, 04510, M\'exico, D.F.; octavio@astroscu.unam.mx}

\begin{abstract} 
We study the phase space available to the local stellar distribution
using a Galactic potential consistent with several recent observational
constraints. We find that the induced phase space
structure has several observable consequences. The spiral arm contribution to the kinematic structure in the solar neighborhood may be as important as the one produced by the Galactic bar. We suggest that some
of the stellar kinematic groups in the solar neighborhood, like the
Hercules structure and the kinematic branches, can be
created by the dynamical resonances of self-gravitating spiral arms and
not exclusively by the Galactic bar.  
A structure coincident with the
Arcturus kinematic group is developed when a hot stellar disk population is considered, which introduces a new perspective on the
interpretation of its extragalactic origin. 
A bar-related resonant mechanism can modify this kinematic structure.
We show that particles in the dark matter disk-like structure predicted by recent LCDM galaxy formation experiments, with similar kinematics to the thick disk, are affected by the same resonances, developing phase space structures or dark
kinematic groups that are independent of the Galaxy assembly history
and substructure abundance. 
We discuss the possibility of using the stellar phase space groups as constraints to non-axisymmetric models of the Milky Way structure.

 \end{abstract}                
 
 \keywords{dark matter --- Galaxy: disk  --- Galaxy: evolution --- Galaxy: kinematics and dynamics --- Galaxy: structure --- solar neighborhood}

\section{Introduction}                                                     
\label{sec:intro}     

The study of stellar kinematic groups in the solar neighborhood has a
long tradition in Galactic astronomy, going back to the
discovery of the Hyades and Ursa Major groups \citep{1869Proct}. In
\citet{antoja08} we presented a study of the solar neighborhood
kinematic groups using a sample of 24,190 stars. We confirmed the
existence of the Sirius, Coma Berenices, Hyades-Pleiades and Hercules
branches (Fig.~\ref{fig}a).  They all have a negative slope of
$\sim16\deg$ in the $U$--$V$ plane (following the standard definitions of $U$ and $V$) and a slight curvature.

Considerable amount of work has been performed in an attempt to explain the origin of
these kinematic groups. External processes like past accretion events
\citep{Navarro04} and internal disk processes like star formation
bursts or secular dynamics have been proposed. Although they were initially considered mutually exclusive, all such mechanisms are
natural in current galaxy formation models
\citep{klypin2008,romano08,CeverinoGF}. Recently, the hypothesis of the disk-dynamical origin of these structures has gained popularity partially because of the
consistency of the Hercules structure with the effects of the Galactic
bar resonances \citep{kalnajs91,dehnen00,fux01}. In addition, steady
or transient spiral arms were proposed to explain the characteristics
of some of these stellar groups
\citep{skuljan99,desimone04,famaey05,quillen05}. 
A more recent study
argues that the combined effect of a bar and spiral
arms is
necessary to accurately reproduce observations in the solar
neighborhood \citep{chakrabarty07}. In that case, spiral arms at the weaker end of the observational range
 (with a fractional amplitude of less than 4$\%$ of the background disk density)
were used. However, in that study the spiral arms seem to
contribute only in the fine structure, thus weakening constraints on the Milky Way (MW)
spiral structure based on the solar neighborhood kinematics. 
Previous studies have considered models for the non-axisymmetric components of the Galaxy motivated by the dynamical point of
view, in particular weak spiral arms described by a cosine function
and bars described by the quadrupole
perturbation.  However, it is unclear whether there is any
dependence of the induced local solar neighborhood kinematics on the
detailed Galactic structure. Moreover, the initial conditions hardly
consider the evolution of the MW.

Our contribution improves upon earlier studies. First we investigate the stellar
kinematic response to a model that satisfies published
observational constraints to the MW structure, in particular to the
non-axisymmetric components. Secondly, we consider for the first time initial
conditions and integration times that attempt to represent stars born
at different times and with different kinematic conditions,
like those in the solar neighborhood.
Lastly, we investigate effects on the local dark matter kinematics, in
particular in the disk-like dark matter structure recently predicted
by LCDM models.  This issue is important to predict signals in direct
dark matter detection experiments.

\section{Simulations}
In order to study the effect of the non-axisymmetric Galactic
structure on the solar neighborhood kinematic distribution, we have
performed numerical integrations of test particle orbits on the
Galactic plane,
adopting the initial conditions discussed in Sect. \ref{ic} and the potential
described in Sect. \ref{potential}. Each particle integration time is
initialized at a value $t=-\tau$ and ended at t$=0$, where $\tau$ is the
time during which the particle is exposed to the non-axisymmetric potential. The induced
kinematic distribution at the end of the simulation is studied by considering
the particles inside a circle 
of radius $500 \pc$ centered at
the solar position. Finally, the predicted and the observed
distributions are compared. The motion equations were integrated with the Bulirsch-Stoer algorithm of \citet{press1992},
conserving Jacobi's integral within a relative variation of
$|(E_{Ji}-E_{Jf})/E_{Ji}|\approx 10^{-11}$ for only-arms and only-bar
models. The reference frame used for the calculations is the rotation frame of the spiral arms when only this non-axisymmetric component is considered and the one of the bar in the other cases. In all cases, we check that the number of particles in the final distributions is statistically robust.

\subsection{The Galactic Model}\label{potential}

\begin{deluxetable}{lcc}
\tablecolumns{2}
\tablewidth{0pt}
\tablecaption{Non-axisymmetric Galactic Disk Components \\
\citep{Pichardo03,Pichardo04}}
\tablehead{\colhead{Parameter} &\colhead{Value}}
\startdata
Bar axis ratio                           &10:3.12             \\  
Bar scale lengths ($\kpc$)               &1.7 and 0.54 \\  
Bar angle respect to the Sun ($\deg$)    & 20         \\
Bar mass ($\msun$)                       & $10^{10}$   \\
Bar pattern speed (see text) $\Omega_b$ ($\kmskpc$) & 60, 45\\       
Spiral arm pitch angle  ($\deg$)         & 15.5  \\
Spiral arm scale-length ($\kpc$)  & $2.5$ \\
Spiral arm mean force ratio (\%)              & 10    \\
Spiral arm pattern speed $\Omega_{sp}$ ($\kmskpc$) & 20        
\label{tab:parameters}
\enddata
\end{deluxetable} 

We use the Galactic potential described by \citet{Pichardo03,
Pichardo04} consistent with several recent observational
constraints. For our study the most relevant parts are the bar and
spiral arm components, whose corresponding parameters are
presented in Table \ref{tab:parameters}. The model is compared with observations and with other models elsewhere \citep{Pichardo03, Pichardo04}. The bar consists of a prolate mass distribution
that resembles Model S of \citet{Freudenreich98} from COBE/DIRBE
data. We consider two alternative bar pattern speeds (60 and 45
$\kmskpc$) according to the range found by \citet{Debattista02}. 
The spiral arm model is a 3D steady model with two arms that traces the
locus reported by \citet{drimmel01} using K-band observations, and also in agreement with the latest results from Spitzer (Benjamin 2009, private communication). 
The solar radius is close to the bar outer Lindblad
resonance (OLR) and to 4:1 inner Lindblad resonance (ILR) of the spiral arms. The ratio of the radial
force of the spiral arms to the background peaks at
$\sim$1.5:10 and the mean value along the arm is approximately 1:10 \citep[see][]{Pichardo03}.
 This is at the higher end of the limits suggested by
\citet{Patsis91} for MW type
galaxies. Instead of introducing the spiral arms and bar adiabatically
to avoid transient features on the model, we prefer to test
self-consistency through an analysis of the stellar orbital
reinforcement of the potential, as in \citet{Patsis91}, presented in
\citet{Pichardo03, Pichardo04}.

\subsection{Initial Conditions}\label{ic}

The choice of appropriate initial conditions is somewhat
controversial. Traditionally they are motivated by self-consistency
with the present stage of the Galactic disk structure. However,
current scenarios of galaxy formation and evolution predict that both
external and internal perturbation mechanisms affect the disk
kinematics. The integration time in our simulations is intended to
consider only the latest stages in the Galaxy evolution. This late
evolution was affected or even dominated by secular dynamical
processes \citep{klypin2008}. We avoided longer integration times 
(more than $2\Gyr$)
because they might require the inclusion of Galactic evolution
like bar weakening or transient arms, as well as the effect of
external perturbations to the MW disk, thus masking the effect of the
current MW structure. 
Therefore we focused on the recently induced
kinematic structure in the solar neighborhood. We explored three
different types of initial conditions (IC1, IC2, IC3), all of them
assuming an exponential disk scale length with $R_\rho=2.5 \kpc$.

IC1--- The initial velocity distribution relative to the Regional
 Standard of Rest is adopted as a Gaussian (with
 $\sigma_U=\sigma_V=5\kms$) constant for all radii. Each particle is
 exposed to the non-axisymmetric
 perturbations for a time ($\tau$) chosen at random between $0$ and $2\Gyr$.
 This maximum integration time corresponds approximately to 14-20 and 6 revolutions of the bar and the spiral arms, respectively.  
With this set of initial conditions we aim to simulate the birth of stars in the disk with small velocity dispersion and the effect of having stars with different ages (different integration time) at the solar position\footnote{
N-body disk simulations assuming a low Q parameter \citep[e.g.][]{thomasson91}   show that a configuration with strong spiral arms and low velocity dispersion can be sustained for at least 7 disk revolutions. These experiments are similar to our integrations using IC1. }.

IC2--- 
The distribution function satisfies the
collisionless Boltzmann equation as it is discussed in \cite{hernquist93}. The velocity dispersion has an
exponential profile with scale length $R_\sigma=7.5\kpc$ and local
normalization $\sigma(R_\odot=8.5 \kpc)\sim20 \kms$. In this case the
integration time is fixed at $\tau=400 \Myr$ for all particles (corresponding approximately to 3-4 bar revolutions and similar to the value of \cite{dehnen00}). With these initial conditions, we can study the relatively rapid induced effects of the non-axisymmetric component on the local kinematics.

IC3--- Identical to IC2 but with a higher velocity dispersion of
$\sigma(R_\odot=8.5 \kpc)\sim40 \kms$, closer to an old thin disk
component, or to the thick disk.  Recent studies of the formation of MW mass galaxies predict that a flattened dark matter structure mirroring the thick
disk properties will form in a LCDM Universe \citep{bruch08,read08}. Therefore, these initial
conditions are also consistent with particles in the dark disk.

\section{Spiral Arm and Bar Contribution to the Local Kinematic Structure}  

\paragraph{Models Using Only Spiral Arms}

\begin{figure*}  
\includegraphics[width=0.32\textwidth]{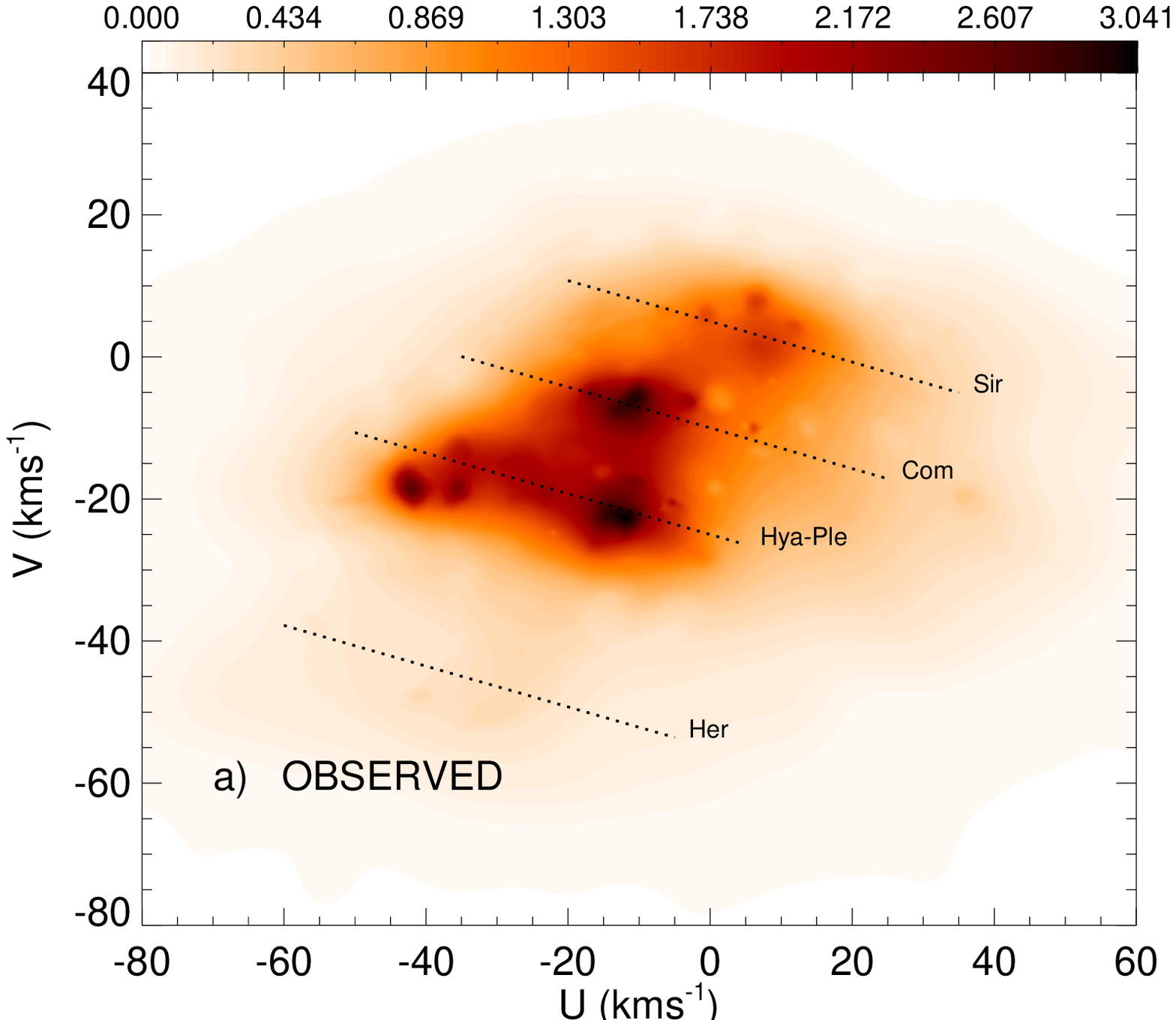} 
\includegraphics[width=0.32\textwidth]{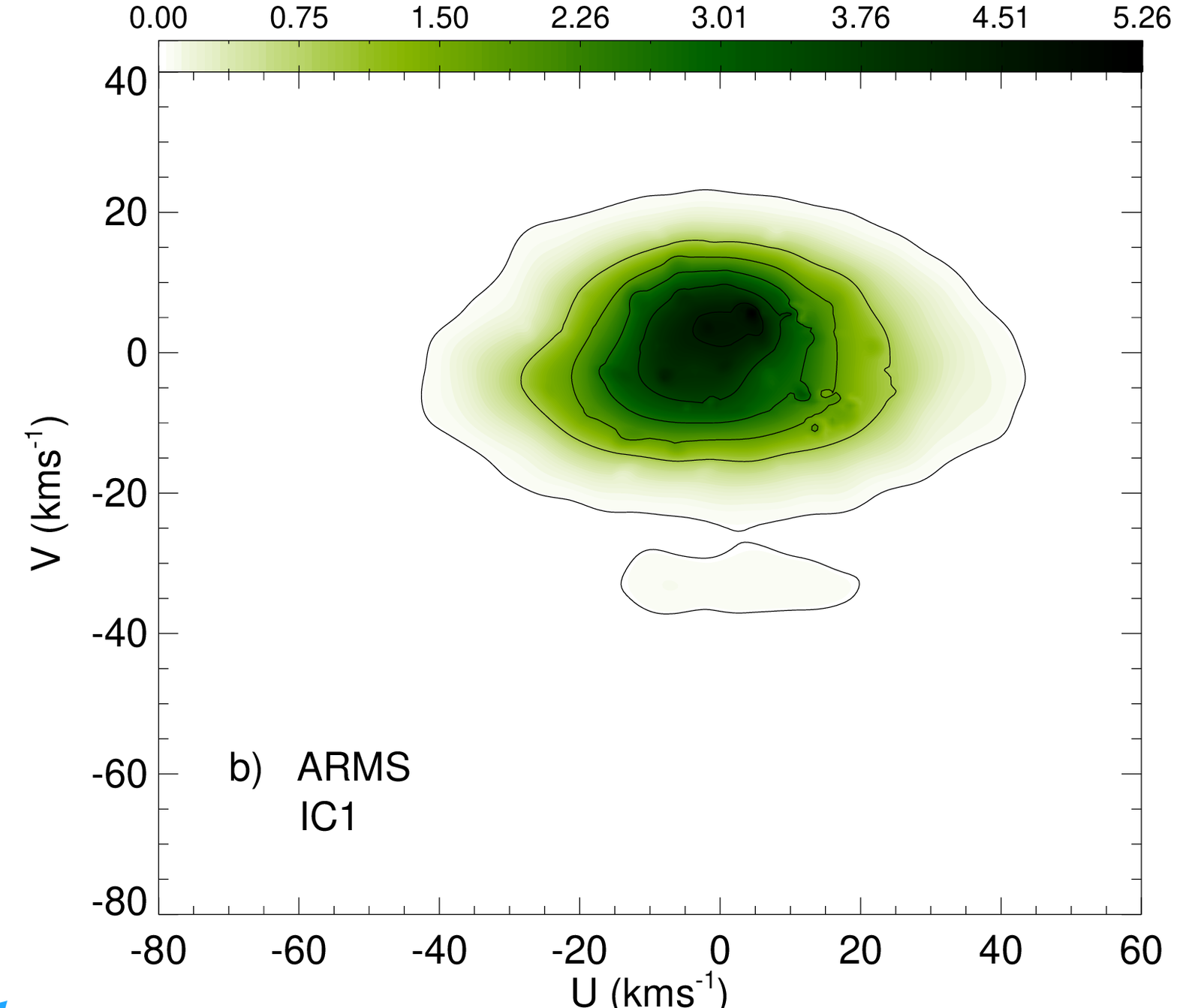}
\includegraphics[width=0.32\textwidth]{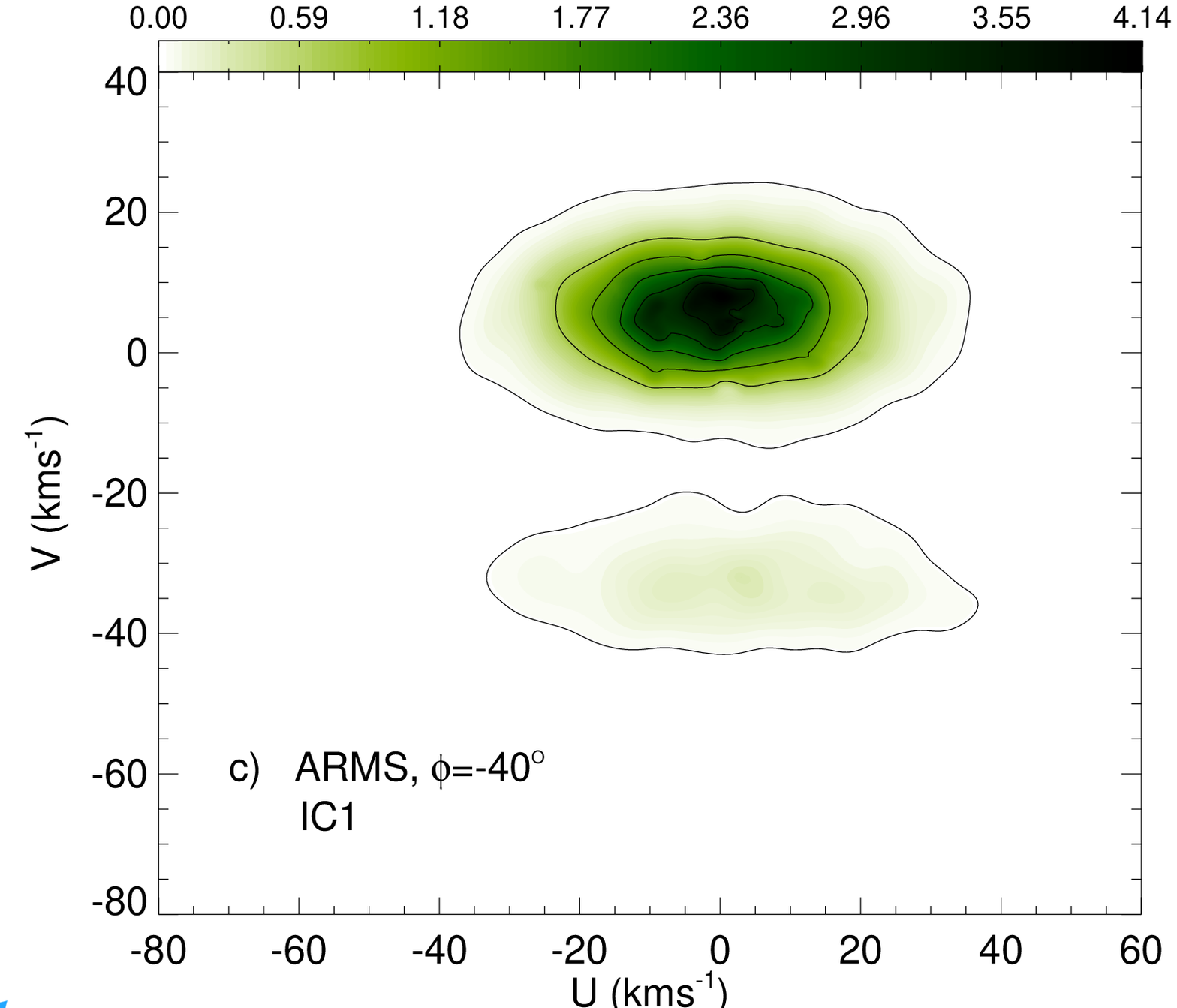}

\includegraphics[width=0.32\textwidth]{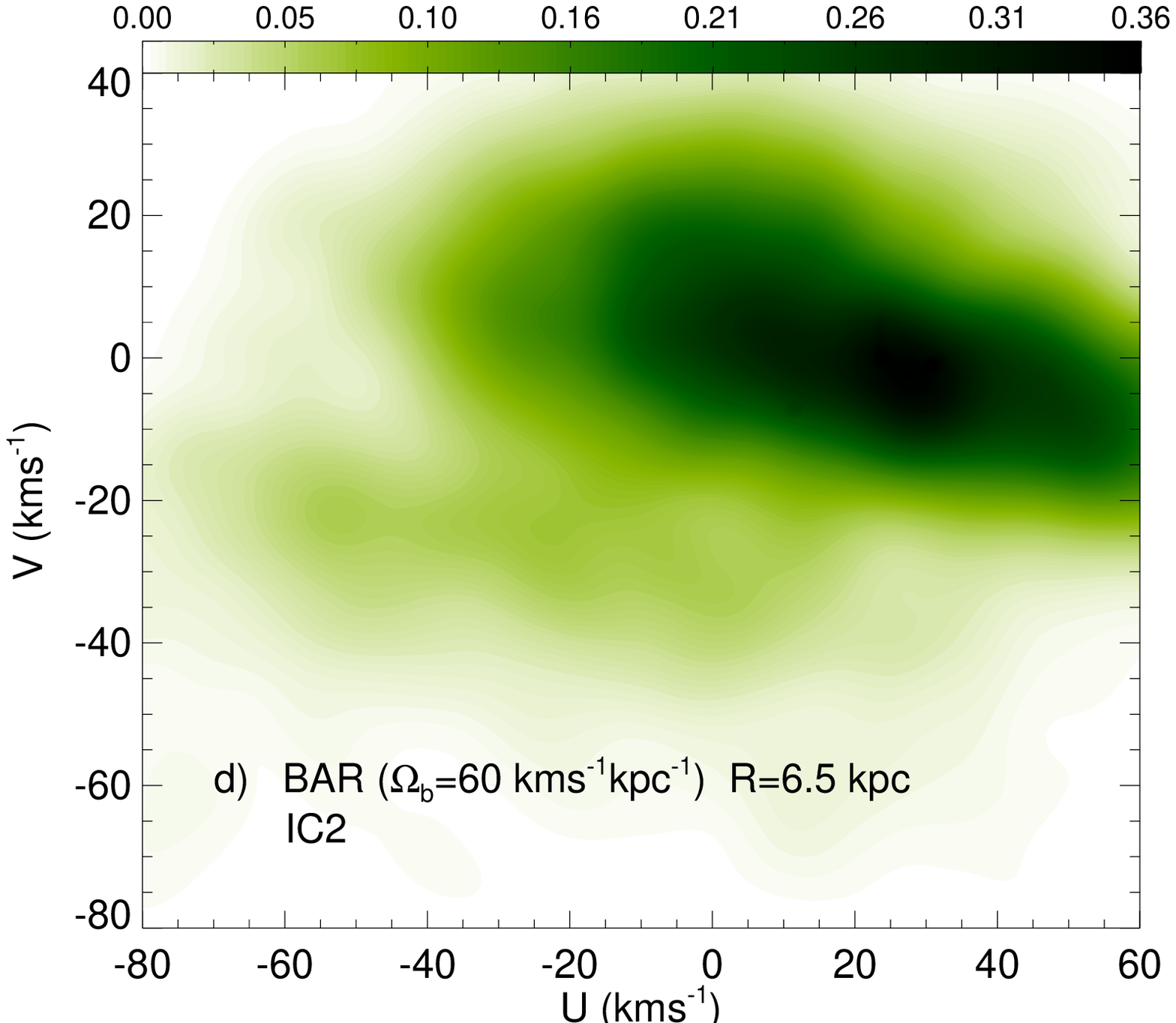}
\includegraphics[width=0.32\textwidth]{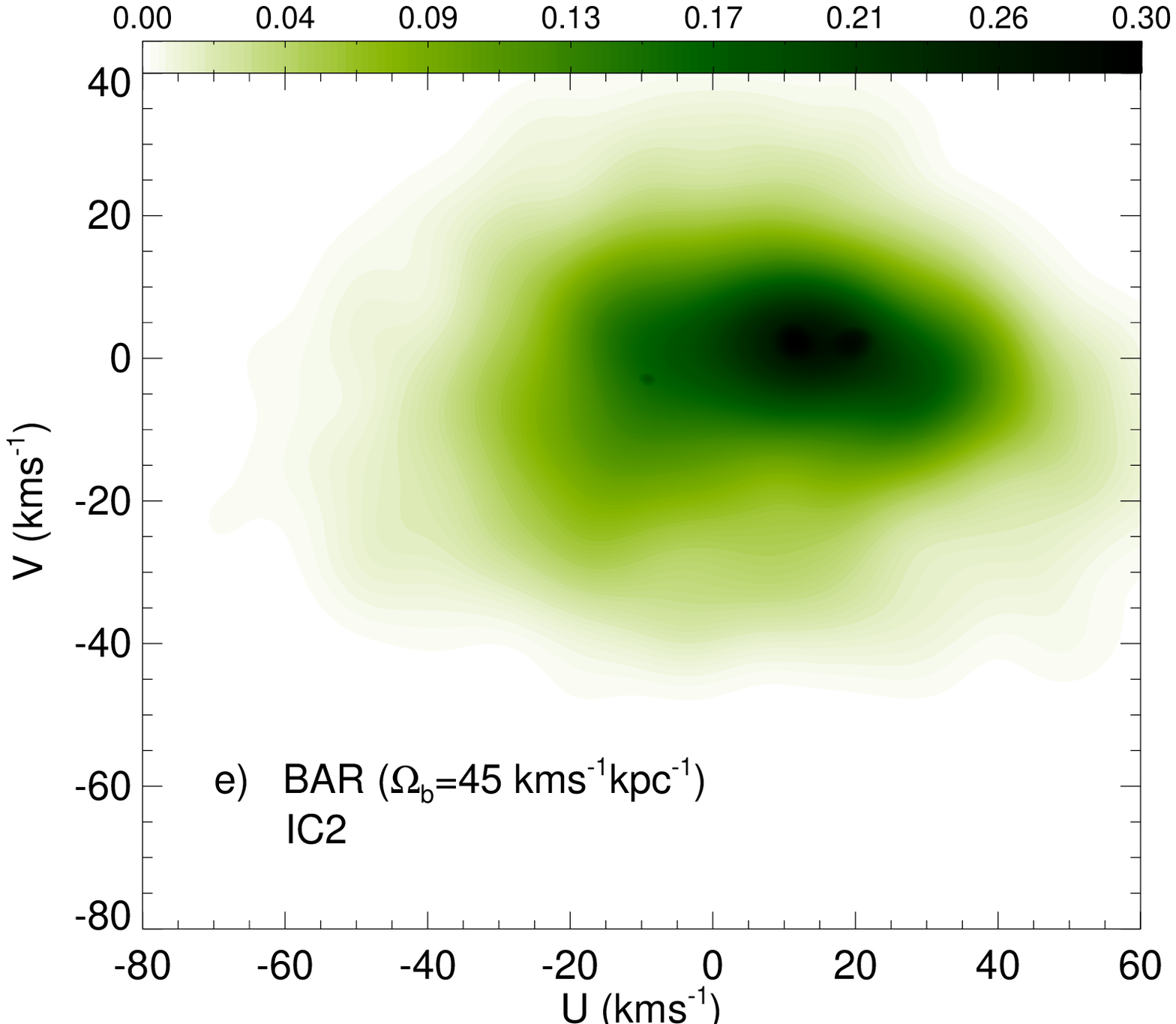}
\includegraphics[width=0.32\textwidth]{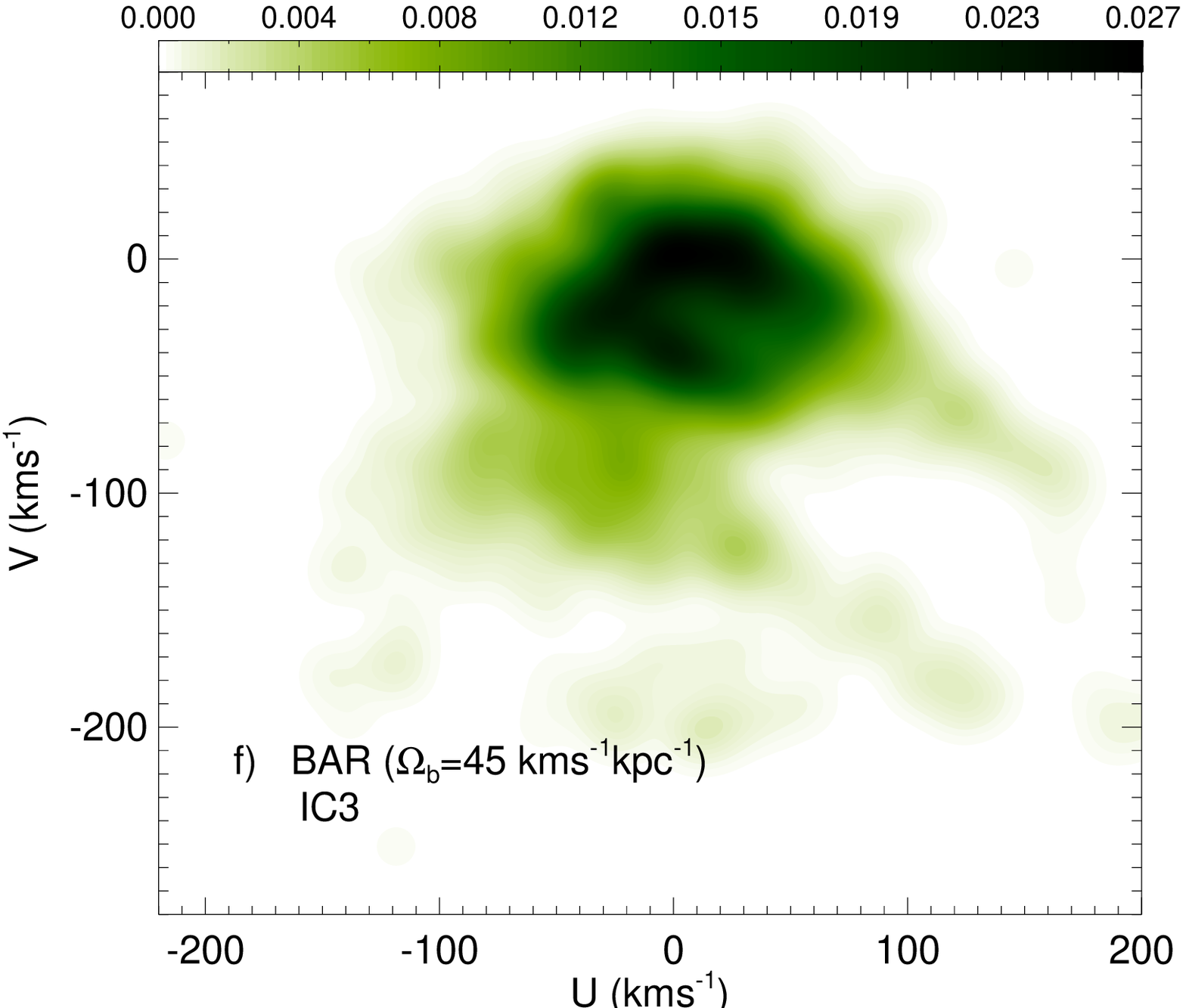}

\includegraphics[width=0.32\textwidth]{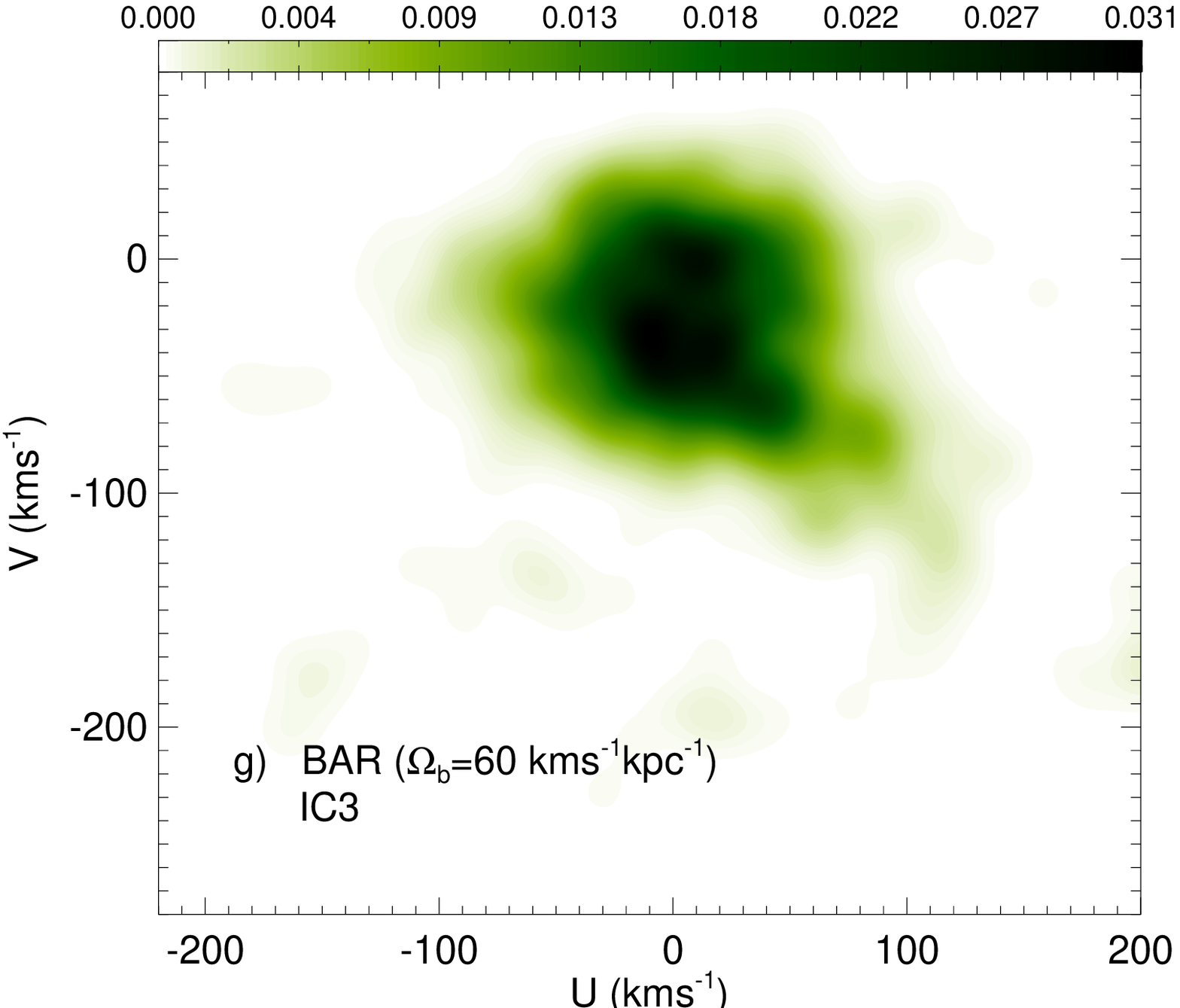}
\includegraphics[width=0.32\textwidth]{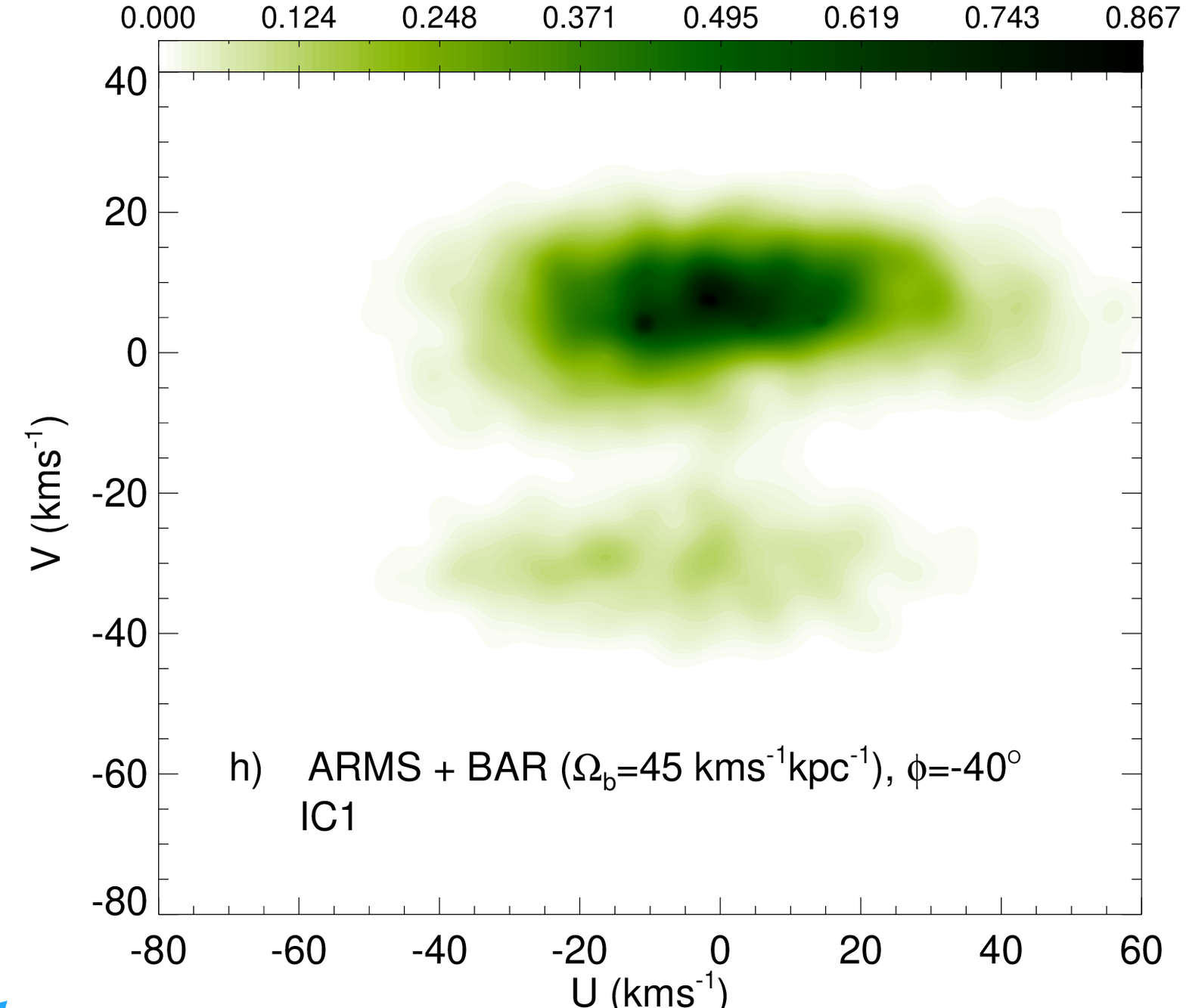}
\includegraphics[width=0.32\textwidth]{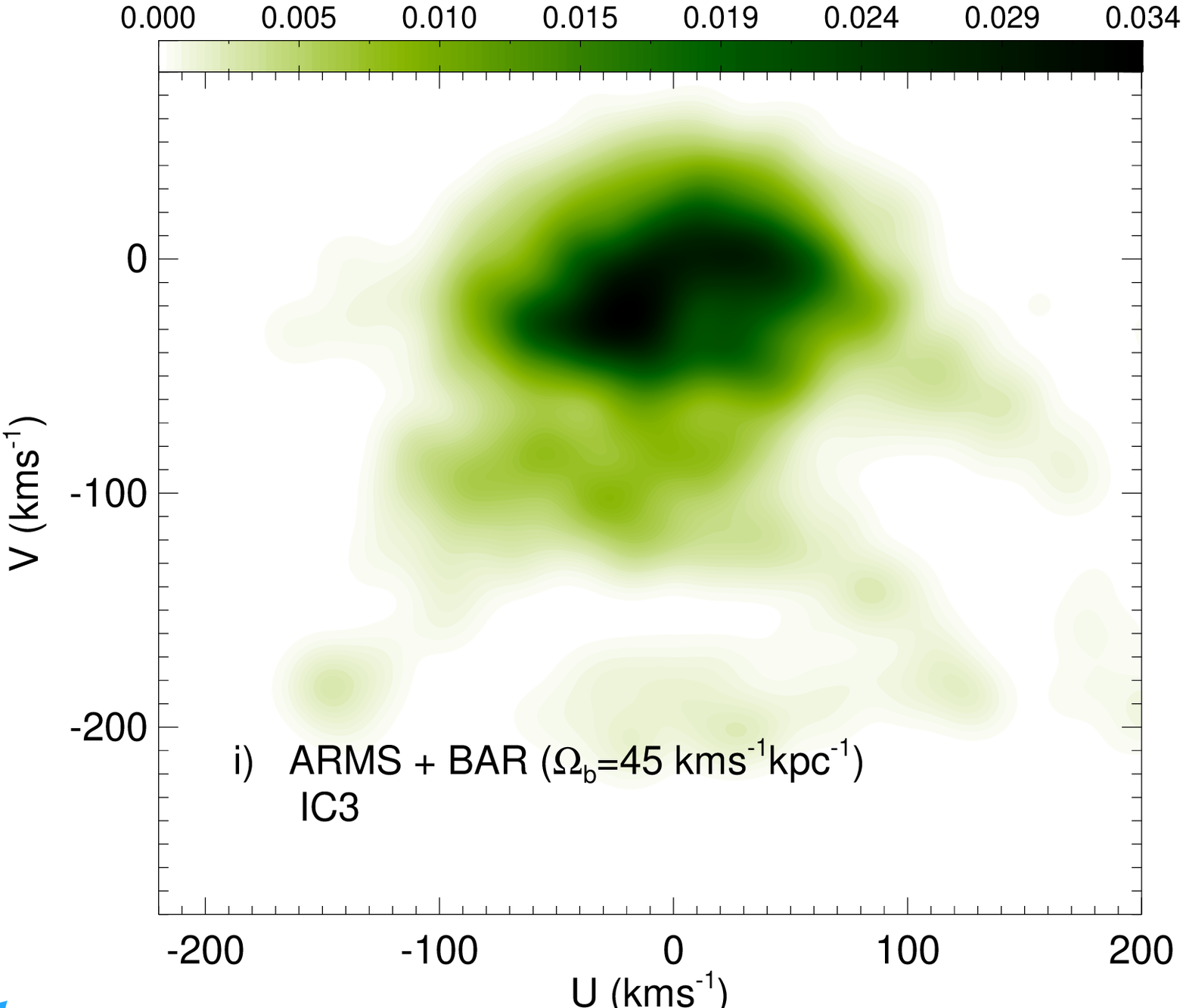}
\caption{Density field in the $U$--$V$ plane obtained by wavelet denoising for: a) the observed sample; b, c) simulations with a model using only spiral arms;  d, e, f, g) simulations with a model using only bar; h, i) simulations with a combined model (bar + spiral arms). Notice the different scales of the panels.}  
\label{fig}
\end{figure*}

In this section we present the kinematic structure developed by the
self-gravitating spiral arm model. We find that this model produces ample substructure. In particular, assuming low velocity
dispersions (IC1), it reproduces a branch at low angular momentum
(with $V\sim-40\kms$ consistent with the Hercules structure) separated
from a central group of substructures. To our knowledge, it
is the first time that an unbarred model has produced a similar
structure. 
Figure~\ref{fig}b shows the $U$--$V$ plane for the solar position. Although the shape of the structure at $V=-40\kms$ is not exactly equal to the Hercules branch, it does show that the spiral arms by themselves crowd the velocity space at these negative $V$. In particular, the test particles in this kinematic group have a perturbation exposure time of at least $\sim 1\Gyr$. The central part of the distribution seems to be split into two groups or branches, resembling some of the observed central kinematic branches (Fig. \ref{fig}a).
In order to explore small variations respect to the configuration of \citet{drimmel01}, given the uncertainty on the exact position of the arms, we also show in Figure~\ref{fig}c the $U$--$V$ plane for a region located at $\phi=-40\deg$, that is at the solar circle but at $40\deg$ of the Sun in the counter-clockwise direction.
In this case and for a wide range of $\phi$, the equivalent to the Hercules structure is also generated. 
We conclude that the contribution of the spiral arms to the solar neighborhood kinematics may be comparable to that of the bar. The sensitivity of our results to the properties of the arms indicates that local kinematics can be used as one of the constraints to the current observational ambiguities about this non-axisymmetric component.

\paragraph{Models Using Only a Bar}

In agreement with previous studies \citep[e.g.][]{dehnen00}, we find
that a Galactic model using only a bar can trigger a kinematic group
at negative $V$ and $U$ resembling the Hercules branch. However, in our
experiments this branch can be populated only under some combinations
of model parameters and initial conditions.  For example, a bar
pattern speed of $\Omega_b=60\kmskpc$ ($R_{OLR}/R=0.74$) under IC2
produces a branch at an incorrect position ($U \geq 0$). However, if
the imaginary solar observer is located at an inner radius $R=6-7\kpc$
(where $R_{OLR}/R=0.97$) a structure appears at $V\sim-35\kms$ and
$U\leq0$, as is shown at the top left of Figure~\ref{fig}d. A
kinematic group at $V$, $U\leq0$ also appears if the bar pattern speed is
reduced to $45\kmskpc$ (Fig.~\ref{fig}e), although the mean
branch velocity $V$ is now slightly smaller: $V\sim-20\kms$ (again
with $R_{OLR}/R=0.97$). In summary, the Hercules branch at the correct
position may be achieved by varying either 
the model pattern speed,
the observer position or even the initial conditions.

If we consider hotter initial conditions (IC3), arch-shaped structures
appear at lower $V$. These structures appear even if we consider an
axisymmetric disk model, as also recently reported by
\citet{minchev09}
who assumed a non-relaxed
model.  In our simulations the positions of these kinematic arches are
modified when the bar is added to the model. This suggests that both
kinematic initial conditions and Galactic structure contribute to create
the kinematic groups.  For example, differences are observed if we use
either $\Omega_b=45\kmskpc$ (Fig.~\ref{fig}f) or
$\Omega_b=60\kmskpc$ (Fig.~\ref{fig}g).  The arches
developed at $V\sim-40\kms$ and $-100\kms$ for the lower bar pattern speed (not seen  when $\Omega_b=60\kmskpc$)
are closer to the Hercules and Arcturus observed
structures. 
\footnote{Notice that the observed Arcturus structure is out of the range of Figure~\ref{fig}a as this structure is not prominent in the sample used by \citet{antoja08}. See Figure 2 of \citet{williams08} for a $U$--$V$ plane containing the Arcturus structure.}
This result should be considered only as evidence of
sensitivity to the model but not necessarily as favoring a
particular $\Omega_b$ value. Furthermore, these simulations show the important role of the bar in the development of the local kinematic
structure.


\paragraph{Combined Models} 

Here we present the results obtained with the model that includes both
the bar and the spiral arm perturbations. 
This model, with different pattern speeds for the bar and the spirals, is supported by recent studies like \citet{patsis09}. 
They reported that the best match between observations and models for the galaxy NGC 3359 is found when different pattern speed for the bar ($39\kmskpc$) and for the spiral arms ($15 \kmskpc$) is considered. 
Now we show examples of how the branches generated by the spiral arms or the bar are maintained when the combined model is used. Figure \ref{fig}h shows the results for the spiral arms plus the bar with $\Omega_b=45\kmskpc$ used with IC1 near the solar position ($\phi=-40\deg$).
In Figure \ref{fig}i, we show the case corresponding to spiral arms and a bar with $\Omega_b=45\kmskpc$ under IC3.  Here the arch-shaped structures associated with Hercules and Arcturus are still observed.

The diversity of the initial conditions in
our simulations is not extensive. Moreover, the observed velocity field is likely to be 
the result of the combined action of several Galactic processes.
Therefore, comparison between the observed velocity distribution\footnotemark[2]
(Fig.~\ref{fig}a) and the results of test particle simulations
is not straightforward. A first approach might be to compare the
observed velocity field with the combined model results under IC1
and IC3, where the central branches and the kinematic groups at low
angular momentum would appear simultaneously.  The next complexity
level might require the inclusion of age and kinematic criteria in the
definition of the artificial observational sample. We defer such
details to a future study, since they do not alter the
conclusions of this Letter.

\section{Local dark matter kinematics}

Another unexpected aspect of the bar- and spiral arm-induced phase
space structure is the effect on the local dark matter kinematics. 
Particles in a possible dark matter bar \citep{Colinetal06}, dark disk
\citep{bruch08,read08} and even in the dark matter halo
\citep{athanassoula05} are trapped/scattered in the same
resonances as stars are.
It has been shown that different bar-induced resonances can be populated by disk-like and halo particles \citep{athanassoula02,athanassoula03,CeverinoResonances}.  

Our study considers only the case of the dark disk.
Following \citep{read09} we assume that our IC3 initial conditions
set also represents the dark particles in the dark thick disk.
Thus Figures \ref{fig}f, \ref{fig}g and \ref{fig}i, obtained from IC3, also reproduce the local dark matter kinematics induced by our MW dynamical models.
Our results show that these models
generate dark matter currents inside the Galactic dark disk.
These dark-matter currents would be independent of the Galactic assembly
history or the dark substructure abundance.
The dynamical history of the Galaxy and its detailed large scale structure may help to establish whether the amplitude of the dark matter kinematic structure is detectable by planned dark matter detection experiments.

%
\section{Discussion and Conclusions}      

The orbital analysis of a MW model consistent with published
observational constraints has several unexpected consequences for
the local stellar kinematics.

The spiral arm contribution to the resonant structure in the solar
neighborhood may be comparable to that of the Galactic bar. The main differences to previous studies are the arm
force contrast and force field shape \citep[Fig. 5, 8 and 9]{Pichardo03}, as well as
the variety in initial conditions. In particular, we find that the
Hercules structure may be produced by the spiral arms
and not exclusively
by bar resonances as traditionally believed. 
\citet{dehnen00} concluded that the Hercules branch is unlikely to have been produced by resonant scattering processes due to spiral arms. His main argument is that spiral arms do not act on stars with
epicycles greater than the interarm separation. However, the epicycle amplitude of the stars in the structure created in our simulations\footnote{This value is obtained theoretically from the epicycle approximation \citep[e.g.][]{asiain99b} or directly from orbital integration.} (at $U\sim[-30,30]$ and $V\sim[-45,-35]$) is about $3\kpc$. On the other hand, the interarm
separation in the different loci of our model is in the range
$5.5-7\kpc$. This is sufficient to produce the Hercules structure and still in agreement
with \citet{dehnen00}.
Other authors like \cite{quillen05}
reproduced the Hyades-Pleiades and Coma Berenices branches
as ascribed to periodic orbits related to the spiral arm 4:1 ILR, but not the Hercules structure.  For some parameter
combinations our model reproduces both kinds of
structure. Detailed comparison with other authors is not straightforward because many of them used the tight-winding approximation to model the spiral arms, a
different simulation strategy, and in some cases, even a four-armed
model \citep{chakrabarty07}, while ours includes two arms.

As in previous studies, we confirm that the Hercules structure can also
be produced by the Galactic bar under certain combinations of the
model parameters and initial conditions.  However, we considered a
prolate bar \citep[for the detailed force field see Fig. 16 of][]{Pichardo04} instead of the widely used cosine bar potential
\citep{dehnen00}. 
The study of possible differences in the local kinematic
structure generated by each model is postponed for a future study.

A subset of our experiments develops a structure resembling the
Arcturus kinematic group. The required condition seems to be a relatively hot stellar disk population similar to the
thick disk in kinematics.  This kinematic group may have arisen from a past accretion event
\citep{Navarro04,villalobos09}.  Alternatively \citet {williams08}
recently discussed a possible disk-dynamical origin of Arcturus based on stellar
population evidence, and postulated the bar 6:1 OLR as the triggering
mechanism. On the other hand, \citet{minchev09} proposed an
origin related to non-equilibrium initial conditions. Our results
support an internal disk origin and, in addition, we find that the dynamics of the bar has an influence on these low angular momentum kinematic groups as it modifies these structures generated in the
axisymmetric model.
Preliminary results of orbit integration in 3-D show that the main arches generated in the 2-D case are maintained and the Galactic bar significantly takes part in defining their shape. However, the 3-D case deserves special attention and is the subject of ongoing studies.
The fact that a vrms of $40\kms$ is high enough to allow the creation of these arches in our experiments
is encouraging, but it is still not sufficient to disentangle the origin of
Arcturus. A deeper discussion of this point is beyond of the scope of
this letter, but it is the subject of an upcoming study.

We show that the Galactic non-axisymmetric potential develops dark kinematic groups  in the dark disk predicted in cosmological simulations of galaxy formation.  These currents are independent of the halo substructure abundance and
may have new consequences for planned dark matter direct detection experiments.      

The dependence of the stellar kinematics in the solar neighborhood  on the
structure, dynamics and initial conditions of our experiments suggests
that kinematic groups may provide a useful constraint on
non-axisymmetric MW models. A systematic scan of the associated
parameter space will be required in order to disentangle the origin of
the different kinematic groups in the solar neighborhood.  We are
currently exploring strategies to do so. Studies like the ones
discussed in this letter will derive benefit from upcoming
surveys like GAIA and SEGUE2. In summary, the imprints of the
non-axisymmetric Galactic structure on the local stellar kinematics
are strong.

\acknowledgments

We thank G. Lake and D. Ceverino for enlightening discussions about
dark matter in the MW, K. Freeman for suggesting the study of
Arcturus in our simulations, and J. Gonz\'alez, V. Avila-Reese,
L. Aguilar, and H. Throop for a careful reading of the manuscript.
 We also thank the referee for the very helpful comments and suggestions which much improved the first version of this paper.
This study was supported by MEC contracts ESP2006-13855-C02-01 and
AYA2006-15623-C02-02, grants CONACyT 60354 and 50720, and UNAM PAPIIT
IN1 19708.  T.A. was supported by the Predoctoral Fellowship of the
Generalitat de Catalunya 2008FIC 00121 and by the LENAC mobility
program.  The simulations were run at the HP CP 4000 cluster
(KanBalam) in the DGSCA/UNAM.


\begin{thebibliography}{DUM}   

\bibitem[Antoja et al.(2008)]{antoja08}Antoja, T., Figueras, F., Fern\'andez, D., Torra, J. 2008, \aap, 490, 135

\bibitem[Asiain et al.(1999)]{asiain99b}
Asiain, R., Figueras, F., Torra, J. 1999, \aap, 350, 434

\bibitem[Athanassoula(2002)]{athanassoula02} Athanassoula, E.\ 2002, 
\apj, 569, 83 

\bibitem[Athanassoula(2003)]{athanassoula03} Athanassoula, E.\ 2003, 
\mnras, 341, 1179 

\bibitem[Athanassoula(2005)]{athanassoula05} Athanassoula, E.\ 2005, 
New York Academy Sciences Annals, 1045, 168 

\bibitem[Bruch et al.(2008)]{bruch08} Bruch, T., Read, J., 
Baudis, L., \& Lake, G.\ 2008, \apj, 696, 920

\bibitem[Ceverino \& Klypin(2007)]{CeverinoResonances} Ceverino, D. \& Klypin, A.\ 2007, \mnras, 379, 1155 

\bibitem[Ceverino \& Klypin(2007b)]{CeverinoGF} Ceverino, D. \& Klypin, A.\ 2007b, \apj, 695, 292

\bibitem[Chakrabarty(2007)]{chakrabarty07} Chakrabarty, D. 2007, \aap, 467, 145

\bibitem[Col{\'{\i}}n et al.(2006)]{Colinetal06} Col{\'{\i}}n, P., Valenzuela, O., \& Klypin, A.\ 2006, \apj, 644, 687 

\bibitem[Debattista et al.(2002)]{Debattista02} Debattista, V.~P., Gerhard, O., \& Sevenster, M.~N.\ 2002, \mnras, 334, 355 

\bibitem[Dehnen(2000)]{dehnen00} Dehnen, W. 2000, \aj, 119, 800

\bibitem[De Simone et al.(2004)]{desimone04}
De Simone, R.S., Wu, X., Tremaine, S. 2004, \mnras, 350, 627 

\bibitem[Drimmel \& Spergel(2001)]{drimmel01} Drimmel, R., Spergel, D.N. 2001, \apj, 556, 181

\bibitem[Famaey et al.(2005)]{famaey05}
Famaey, B., Jorissen, A., Luri, X., et al. 2005, \aap, 430, 165

\bibitem[Freudenreich(1998)]{Freudenreich98} Freudenreich, H.~T.\ 
1998, \apj, 492, 495 

\bibitem[Fux(2001)]{fux01} Fux, R.  2001, \aap, 373, 511
	
\bibitem[Hernquist(1993)]{hernquist93}Hernquist, L. 1993, ApJS, 86, 389

\bibitem[Kalnajs(1991)]{kalnajs91}
Kalnajs, A. J. 1991, {\em Dynamics of Disk Galaxies}, eds. B. Sundelius, 323

\bibitem[Klypin et al.(2008)]{klypin2008} Klypin, A., Valenzuela, 
O., Colin, P., \& Quinn, T.\ 2008, arXiv:0808.3422

\bibitem[Minchev et al.(2009)]{minchev09} Minchev, I., Quillen, 
A.~C., Williams, M., Freeman, K.~C., Nordhaus, J., Siebert, A., 
\& Bienayme, O.\ 2009,  \mnras, 396,  L56 

\bibitem[Navarro et al.(2004)]{Navarro04} Navarro, J.~F., Helmi, 
A., \& Freeman, K.~C.\ 2004, \apjl, 601, L43 

\bibitem[Patsis et 
al.(1991)]{Patsis91} Patsis, P.~A., Contopoulos, G., \& Grosbol, P.\ 1991, \aap, 243, 373 

\bibitem[Patsis et al.(2009)]{patsis09} Patsis, P.~A., Kaufmann, D.~E., Gottesman,S.~T., \& Boonyasait, V.\ 2009, \mnras, 394, 142

\bibitem[Pichardo et al.(2003)]{Pichardo03} Pichardo, B.,
Martos, M., Moreno, E., \& Espresate, J.\ 2003, \apj, 582, 230

\bibitem[Pichardo et al.(2004)]{Pichardo04} Pichardo, B.,
Martos, M., \& Moreno, E.\ 2004, \apj, 609, 144

\bibitem[Press et al.(2004)]{press1992} Press, W. H., Teukolsky,
S. A., Vetterling, W. T., \& Flannery, B. P. 1992, Numerical Recipes in
Fortran 77: {\it The Art of Scientific Computing}, 2nd ed. (Cambridge:
Cambridge Univ. Press)

\bibitem[Proctor(1869)]{1869Proct} Proctor, R.~A.\ 1869, Royal 
Society of London Proceedings Series I, 18, 169 

\bibitem[Quillen \& Minchev(2005)]{quillen05} 
Quillen, A.C., \& Minchev, I. 2005, \apj, 130,576

\bibitem[Read et al.(2008)]{read08} 
Read, J.~I., Lake, G., Agertz, O., \& Debattista, V.~P.\ 2008, \mnras, 389, 1041 

\bibitem[Read et al.(2009)]{read09} Read, J.~I., Mayer, L., 
Brooks, A.~M., Governato, F., \& Lake, G.\ 2009, arXiv:0902.0009

\bibitem[Romano-D{\'{\i}}az et al.(2008)]{romano08} 
Romano-D{\'{\i}}az, E., Shlosman, I., Heller, C., \& Hoffman, Y.\ 2008, \apjl, 687, L13 

\bibitem[Skuljan et al.(1999)]{skuljan99} 
Skuljan, J., Hearnshaw, J. B., Cottrell, P. L. 1999, \mnras,  308, 731

\bibitem[Thomasson et al.(1991)]{thomasson91} 
Thomasson, M., Donner, K.J., Elmegreen, B.G. 1991, \aap, 250, 316

\bibitem[Villalobos \& Helmi(2009)]{villalobos09} Villalobos, A., \& Helmi, A.\ 2009, arXiv:0902.1624

\bibitem[Williams et al.(2008)]{williams08} Williams, M.E.K., Freeman, K.C., Helmi, A. 2008, arXiv:0810.2669

\end{thebibliography}
\end{document}